 \documentclass[final,5p,times,twocolumn]{elsarticle}

\usepackage{amssymb}
\journal{Physics Letters A}

\begin{document}

\begin{frontmatter}



\title{Convergence towards asymptotic state in 1-D mappings: a scaling
investigation}


\author{Rivania M. N. Teixeira$^{\rm 1}$}
\ead{rivania@fisica.ufc.br, Phone:+55 19 3526 9174}
\author{Danilo S. Rando$^{\rm 2}$}
\author{Felipe C. Geraldo$^{\rm 3}$}
\author{R. N. Costa Filho$^{\rm 1}$}
\author{Juliano A. de Oliveira$^{\rm 2,3}$}
\author{Edson D. Leonel$^{\rm 2}$}
%\ead{edleonel@rc.unesp.br}
\address{$^{\rm 1}$Departamento de F\'isica, UFC - Univ. Federal do Cear\'a,
Fortaleza - Cear\'a - Brazil\\
$^{\rm 2}$Departamento de F\'isica, UNESP - Univ. Estadual Paulista, Av.24A,
1515 - Bela Vista - 13506-900 - Rio Claro - SP - Brazil\\
$^{\rm 3}$UNESP - Univ Estadual Paulista - S\~ao Jo\~ao da Boa Vista - SP -
Brazil}

%\pacs{05.45.Pq, 05.45.Tp}

\begin{abstract}
Decay to asymptotic steady state in one-dimensional logistic-like mappings is
characterized by considering a phenomenological description supported by
numerical simulations and confirmed by a theoretical description. As the
control parameter is varied bifurcations in the fixed points appear. We
verified at the bifurcation point in both; the transcritical, pitchfork and
period-doubling bifurcations, that the decay for the stationary point is
characterized via a homogeneous function with three critical exponents depending on the nonlinearity of the mapping. Near the bifurcation the decay to
the fixed point is exponential with a relaxation time given by a power law
whose slope is independent of the nonlinearity. The formalism is general and
can be extended to other dissipative mappings.
\end{abstract}

\begin{keyword}
Chaos; Survival probability, parameter space, Shrimps.

\end{keyword}

\end{frontmatter}

\date{today}

% \linenumbers


\section{Introduction}
\label{sec1}

Mappings are often used to characterize the evolution of dynamical systems by
using the so called discrete time \cite{Ref0}. The interest in the subject was
increased after the investigation of May \cite{Ref1} with direct application
to biology \cite{Ref2}. After that a large number of applications involving
mappings were considered particularly related to physics
\cite{Ref3,Ref4,Ref5}, chemistry, biology, engineering, mathematics and many
others
\cite{Ref6,Ref7,Ref8,Ref9,Ref10,Ref11,Ref12,Ref13,Ref14,Ref15,Ref16,Ref17}.

The type of dynamics certainly depends on the control parameter. As it is
varied, bifurcations appear changing the dynamics of the steady
state \cite{Ref0}, and for specific rates eventually leads to chaos
\cite{Ref18,Ref19}. Collisions of stable and unstable manifolds yield the
destruction of chaotic attractors \cite{Ref3,Ref20}. Many of these dynamical
properties are already known and are taken indeed at the asymptotic state. The
way the system goes to equilibrium is generally disregarded just by
considering the evolution for a large transient.

Our main goal in this Letter is to apply a scaling formalism to explore the
evolution towards the equilibrium near three types of bifurcations in a
logistic-like mapping: (a) transcritical; (b) pitchfork and (c)
period-doubling. Indeed at the bifurcation point the orbit relaxes to 
equilibrium in a way described by a homogeneous function with
well defined critical exponents \cite{Add1,Add2,Add3}. Such exponents
are not universal and depend mostly on the nonlinearity of the mapping and on
the type of bifurcation. Near a bifurcation, the relaxation to the
equilibrium is exponential, with a relaxation time characterized by a power law
\cite{Add1}. Here, two different procedures are used  to obtain the exponents. The first one is mostly
phenomenological with scaling hypotheses ending up with a scaling law of the
three critical exponents. The second one considers transforming the difference
equation into a differential one and solving it with the convenient initial
conditions. Our analytical results confirm remarkably well the numerical data
obtained via computer simulation.

This piece of work is organized as follows. First the mapping, the equilibrium conditions and a phenomenological approach leading to the scaling law is described. Then we discuss the critical exponents by transforming the difference equation into a differential equation. Moving on we present discussions and extensions to other bifurcation when finally our
conclusions are drawn.

%***********************************************************
\section{The mapping and phenomenological properties of the steady state}
\label{sec2}
\begin{figure}[t]
%\vspace*{-0.8cm}
\centerline{\includegraphics[width=1.0\linewidth]{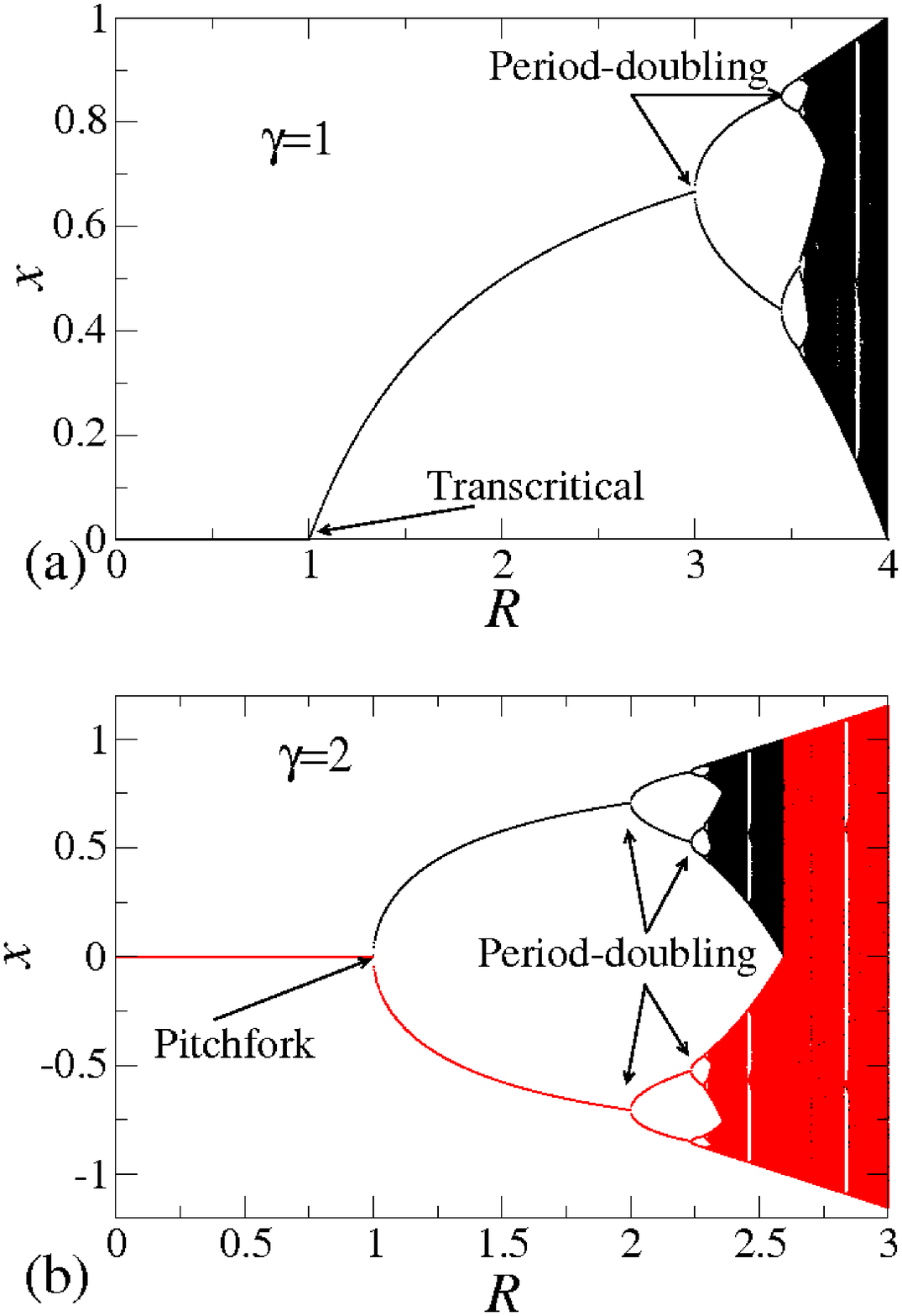}}
\caption{Orbit diagrams obtained for Eq. (\ref{Eq1}) considering: (a)
$\gamma=1$ (traditional logistic map) and; (b) $\gamma=2$ (cubic map). Some
bifurcations are indicated in the figures.}
\label{Fig1}
\end{figure}

The mapping we consider is written as
\begin{equation}
x_{n+1}=Rx_n(1-x_n^{\gamma})~,
\label{Eq1}
\end{equation}
where $\gamma\ge 1$, $R$ is a control parameter and $x$ is a dynamical
variable. A typical orbit diagram is shown in Fig. \ref{Fig1} for: (a)
$\gamma=1$ (logistic map) and; (b) $\gamma=2$ (cubic map).

The fixed points are obtained by solving $x_{n+1}=x_n=x^*$ and two cases must
be considered: (i) $\gamma$ is an even number or; (ii) $\gamma$ is any other
value (odd, irrational etc). For case (i) there are three fixed points. One is
$x^*=0$, which is stable (asymptotically stable) for $R\in[0,1)$ and the two
others are $x^*=\pm [1-1/R]^{1/\gamma}$, which are stable for
$R\in(1,(2+\gamma)/\gamma)$. The bifurcation at $R=1$ is called pitchfork
\cite{Ref21,Ref22}. For case (ii) there are only two fixed points. One is
$x^*=0$, stable for $R\in[0,1)$ and the other is $x^*=[1-1/R]^{1/\gamma}$,
being stable for $R\in(1,(2+\gamma)/\gamma)$. Transcritical is the bifurcation
at $R=1$ for this case. Both bifurcations are identified in Fig.
\ref{Fig1}(a,b).
%As soon as the parameter $R$ is increased a sequence of
%period doubling bifurcation is observed therefore reaching chaos at the
%Feigenbaum rate \cite{Ref18,Ref19}.
Our goal is to consider the convergence to the fixed point $x^*=0$ at the
bifurcation in $R_c=1$ and in its neighbouring such that $\mu=R_c-R\cong 0$,
with $R\le R_c$.

The orbit diagram allows also to extract more properties. After a
transcritical bifurcation, as seen in Fig. \ref{Fig1}(a), the period-1 orbit
is stable for the range $R\in(1,3)$ when a period-doubling bifurcation
happens. After that a period-doubling sequence is observed, obeying a
Feigenbaum scaling \cite{Ref18,Ref19} until reach the chaos. Similar
dynamics, for a different range of $R$ is also observed after a pitchfork
bifurcation, as shown in Fig. \ref{Fig1}(b).

\begin{figure}[t]
%\vspace*{-0.8cm}
\centerline{\includegraphics[width=1.0\linewidth]{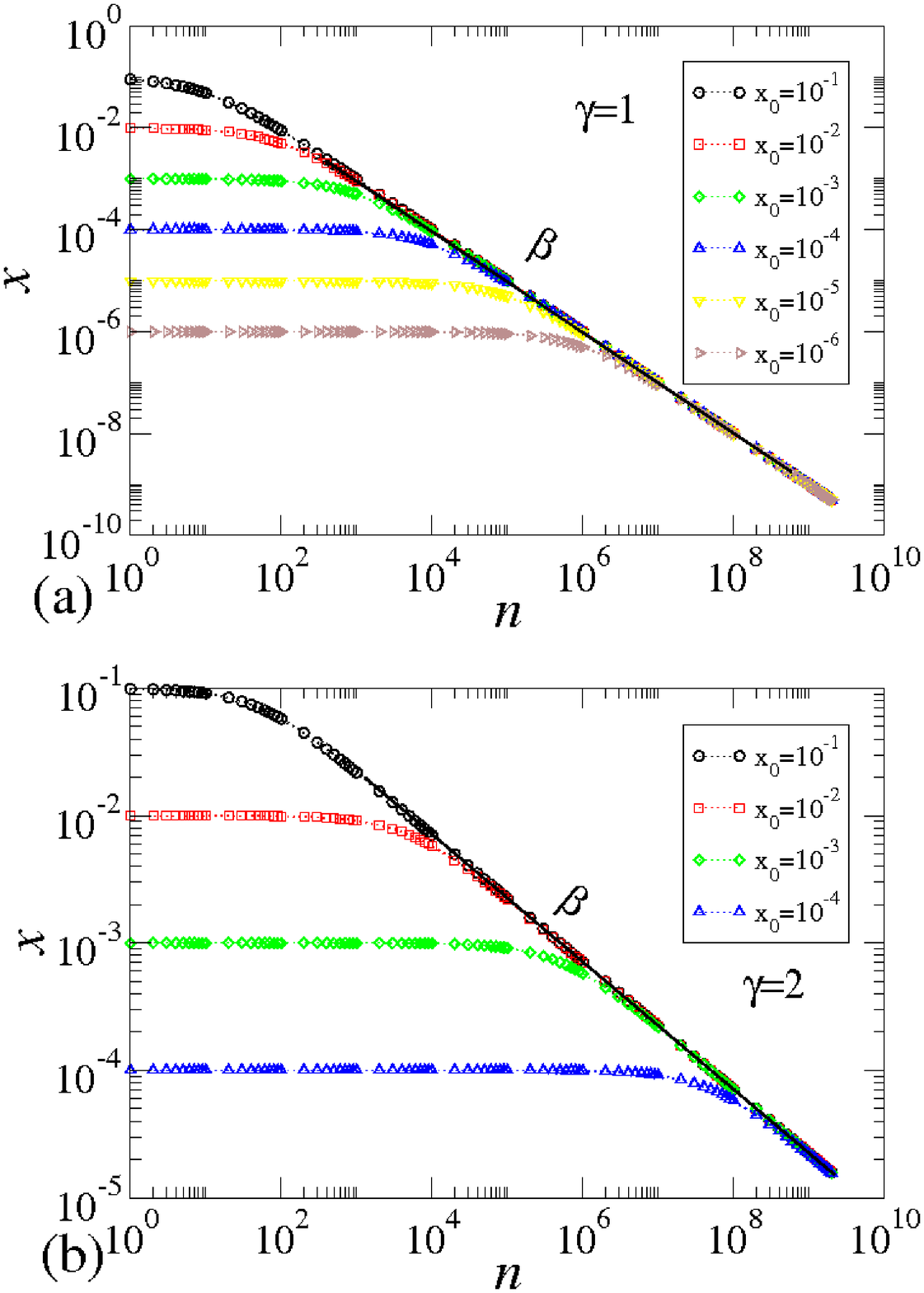}}
\caption{Convergence towards the steady state at $x^*=0$ for: (a) $\gamma=1$
and; (b) $\gamma=2$. The initial conditions are characterized in the figure.
For
this figure, we used the parameter $\mu=0$.}
\label{Fig2}
\end{figure}

The natural variable to describe the convergence to the steady state is the
distance from the fixed point \cite{Add2}. Indeed for the fixed point $x^*=0$,
the distance to the fixed point is the own dynamical variable $x$. The
convergence to the steady state must also depends on the number of iterations
$n$, on the initial condition $x_0$, and of course on the parameter
$\mu=R_c-R$. The parameter $\mu=0$ defines the bifurcation point and the
convergence to the fixed point is shown in Fig. \ref{Fig2} for two different
values of $\gamma$: (a) $\gamma=1$ and; (b) $\gamma=2$ and different initial
conditions $x_0$, as labelled in the figure.

We see from Fig. \ref{Fig2} that depending on the initial condition $x_0$,
the orbit stays confined in a plateau of constant $x$ and, after reaching a
crossover iteration number, $n_x$, the orbit suffers a changeover from a
constant regime to a power law decay marked by a critical exponent $\beta$.
The length of the plateau also depends on the initial $x_0$. Based on the
behaviour observed from Fig. \ref{Fig2} we can suppose that:
\begin{enumerate}
\item
{
For a sufficient short $n$, say $n\ll n_x$, the behaviour of $x~vs.~n$ is
given by
\begin{equation}
x(n)\propto x_0^{\alpha}~,~{\rm for}~n\ll n_x~,
\label{hip1}
\end{equation}
and because $x\propto x_0$, we conclude that the critical exponent $\alpha=1$.
}
\item
{
For sufficient large $n$, i.e., $n\gg n_x$, the dynamical variable is
described as 
\begin{equation}
x(n)\propto n^{\beta}~,~{\rm for}~n\gg n_x~,
\label{hip2}
\end{equation} 
where the exponent $\beta$ is called a decay exponent. The numerical value is
not universal and depends on the nonlinearity of the mapping.
} 
\item
{
Finally, the crossover iteration number $n_x$ is given by 
\begin{equation}
n_x\propto x_0^z~,
\end{equation}
where $z$ is a changeover exponent.
}
\end{enumerate} 

\begin{figure}[t]
%\vspace*{-0.8cm}
\centerline{\includegraphics[width=1.0\linewidth]{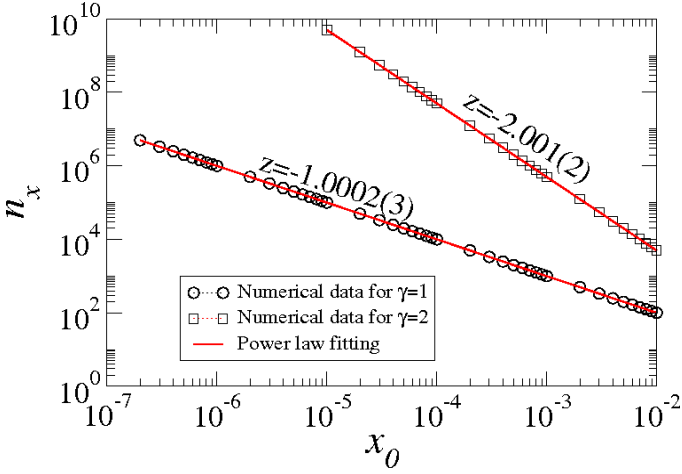}}
\caption{Plot of the crossover iteration number $n_x$ against the
initial condition $x_0$ together with their power law fitting for $\gamma=1$
with slope $z=-1.0002(3)$ and $\gamma=2$ with slope $z=-2.001(2)$.}
\label{Fig3}
\end{figure}

The exponents $\beta$ and $z$ can be obtained by considering specific plots.
After the constant plateau, a power law fitting furnishes $\beta$. Indeed for
$\gamma=1$ (logistic map) we found $\beta=-0.99981(3)$ while for $\gamma=2$
(cubic map) we obtained $\beta=-0.49969(5)$. To obtain the exponent $z$ we
must have the behaviour of $n_x~vs.~x_0$, where $n_x$ is obtained as the
crossing of the constant plateau by the power law decay, as shown in Fig.
\ref{Fig3}. 

The slope obtained for $\gamma=1$, as shown in Fig. \ref{Fig3} is
$z=-1.0002(3)$ while for $\gamma=2$ the exponent obtained is $z=-2.001(2)$.

The behaviour shown in Fig. \ref{Fig2} together with the three scaling
hypotheses allow us to describe the behaviour of $x$ as an homogeneous
function of the variables $n$ and $x_0$, when $\mu=0$, of the type
\begin{equation}
x(x_0,n)=lx(l^ax_0,l^bn)~,
\label{Eq_hom}
\end{equation}
where $l$ is a scaling factor, $a$ and $b$ are characteristic exponents.
Because $l$ is a scaling factor we choose $l^ax_0=1$, leading to
$l=x_0^{-1/a}$. Substituting this expression in Eq. (\ref{Eq_hom}) we obtain 
\begin{equation}
x(x_0,n)=x_0^{-1/a}x(1,x_0^{-b/a}n)~.
\label{Eq_hom1}
\end{equation}
Assuming the term $x(1,x_0^{-b/a}n)$ is constant for $n\ll n_x$ and comparing
Eq. (\ref{Eq_hom1}) with the first scaling hypothesis we conclude that
$\alpha=-1/a$. Moving on and choosing $l^bn=1$, which leads to $l=n^{-1/b}$
and substituting in Eq. (\ref{Eq_hom}) we obtain
\begin{equation}
x(x_0,n)=n^{-1/b}x(n^{-a/b}x_0,1)~.
\label{Eq_hom2}
\end{equation}
Again we suppose the term $x(n^{-a/b}x_0,1)$ is constant for $n\gg n_x$.
Comparing then with second scaling hypothesis we end up with $\beta=-1/b$.
Finally we compare the two expressions obtained for the scaling factor. It
indeed leads to $n_x=x_0^{\alpha/\beta}$. A comparison with third scaling
hypothesis allow us to obtain a relation between the three critical exponents
$\alpha$, $\beta$ and $z$ therefore converging to the following scaling law
\begin{equation}
z={{\alpha}\over{\beta}}~.
\label{law}
\end{equation}

\begin{figure}[t]
%\vspace*{-0.8cm}
\centerline{\includegraphics[width=1.0\linewidth]{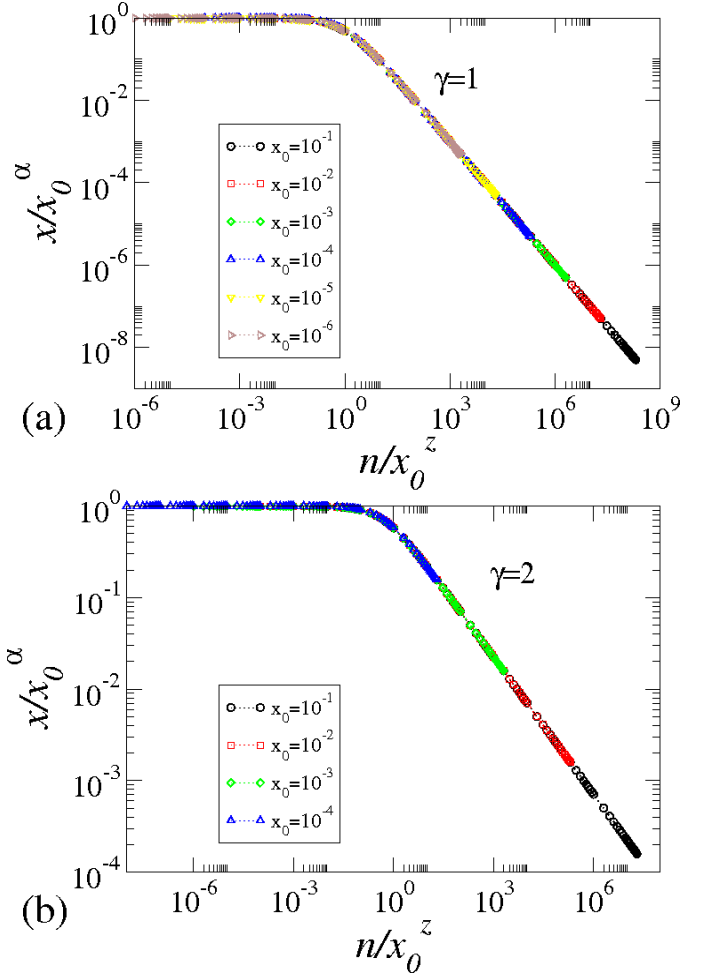}}
\caption{Overlap of all curves shown in Fig. \ref{Fig2} onto a single and
universal plot, after a convenient rescale of the axis, for both: (a)
$\gamma=1$ and (b) $\gamma=2$.}
\label{overlap}
\end{figure}

The knowledge of any two exponents allow to find the third one by using Eq.
(\ref{law}). Moreover the exponents can also be used to rescale the variables
$x$ and $n$ in a convenient way such that $x\rightarrow x/x_0^{\alpha}$ and
$n\rightarrow n/x_0^z$ and overlap all curves of $x~vs.~n$ onto a single and
hence universal curve, as shown in Fig. \ref{overlap}.

\begin{figure}[t]
%\vspace*{-0.8cm}
\centerline{\includegraphics[width=1.0\linewidth]{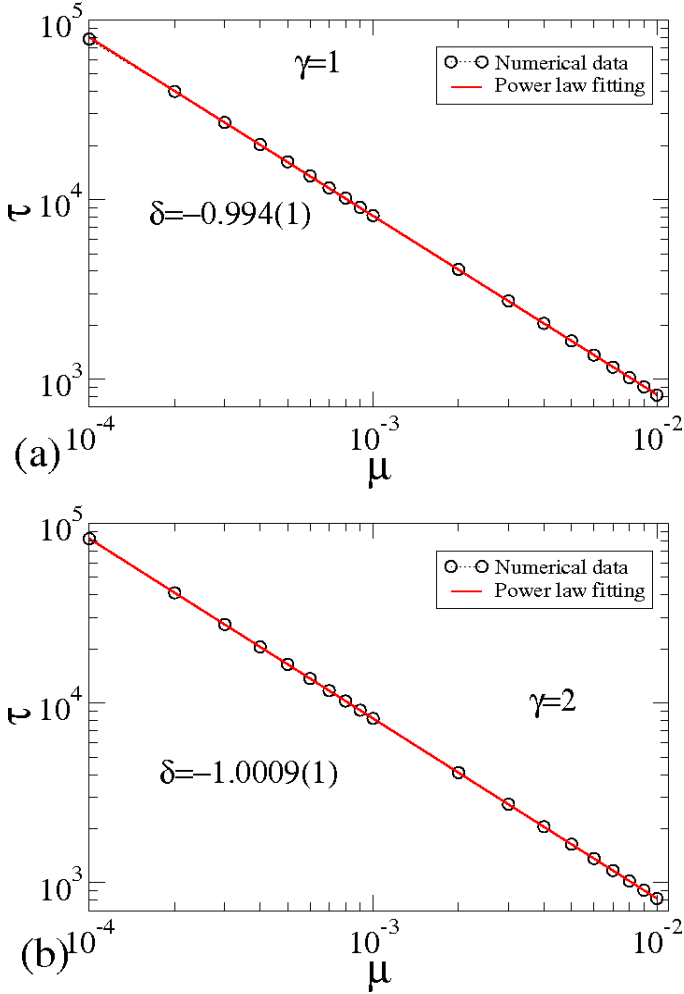}}
\caption{Plot of the relaxation to the fixed point as a function of $\mu$ in
the logistic-like map for the exponents: (a) $\gamma=1$ and; (b) $\gamma=2$.}
\label{Fig4}
\end{figure}

Before moving to the next section and consider the dynamics in a more
analytical way, let us discuss here the dynamics for $\mu\ne0$. This
characterises the neighbouring of the bifurcation. The convergence to the
steady state is marked by an exponential law of the type (see Refs.
\cite{Add1,Add2})
\begin{equation}
x(n,\mu)\propto e^{-n/\tau}~,
\label{decay_exp}
\end{equation} 
where $\tau$ is the relaxation time described by
\begin{equation}
\tau\propto \mu^{\delta}~,
\label{decay_tau}
\end{equation}
where $\delta$ is a relaxation exponent. Figure \ref{Fig4} shows the behaviour
of $\tau~vs.~\mu$ for two different values of $\gamma$.

A power law fitting furnishes the exponent $\delta\cong -1$ and is
independent on the value of the parameter $\gamma$. In the next section we
describe how to obtain the exponents discussed in this section using an
analytical approach.
%**********************************************************

%************************************************************
\section{An analytical description to the equilibrium}
\label{sec3}

Let us now discuss a different approach to reach the equilibrium. We start
first with case (i), i.e., at the bifurcation point $R=R_c=1$. The equation of
the mapping is then written as 
\begin{equation}
x_{n+1}=x_n-x_n^{\gamma+1}~.
\label{a1}
\end{equation}

Very near the fixed point, we suppose the dynamical variable $x$ can be
considered as a continuous variable. Therefore Eq. (\ref{a1}) is rewritten in
a convenient way as (see also Ref. \cite{liv} for a recent application in a
2-D mapping)
\begin{eqnarray}
x_{n+1}-x_n&=&{{x_{n+1}-x_n}\over{(n+1)-n}}~,\nonumber\\
&\approx&{{dx}\over{dn}}=-x^{\gamma+1}~.
\label{a2}
\end{eqnarray}

Grouping the terms properly we obtain the following differential equation
\begin{equation}
{{dx}\over{x^{\gamma+1}}}=-dn~.
\label{a3}
\end{equation}
Indeed the initial condition $x_0$ is defined for $n=0$. Of course for a
generic $n$ we have $x(n)$. Using these terms as limit of the integrals we end
up with
\begin{equation}
\int_{x_0}^{x(n)}{{dx}\over{x^{\gamma+1}}}=-\int_0^n dn~.
\label{a4}
\end{equation}
After integrating Eq. (\ref{a4}) and organising the terms properly we obtain
the following expression
\begin{equation}
x(n)={{x_0}\over{\left[x_0^{\gamma}\gamma n+1\right]^{1/\gamma}}}~.
\label{a5}
\end{equation}

Let us now discuss the implications  of Eq. (\ref{a5}) for specific ranges of
$n$. We start with the case $x_0^{\gamma}\gamma n\ll 1$, which is equivalent
to the previous section of $n\ll n_x$. For such a case we obtain that
$x(n)\cong x_0$. A quick comparison with first scaling hypothesis allow us to
conclude that the critical exponent $\alpha=1$. Second we consider the
situation $x_0^{\gamma}\gamma n\gg 1$, corresponding to $n\gg n_x$ in the
previous section. For such case we obtain that
\begin{equation}
x(n)\approx n^{-1/\gamma}~.
\label{a6}
\end{equation}

Comparing then this result with scaling hypothesis two of the previous section
we conclude $\beta=-1/\gamma$. The last case is obtained when
$x_0^{\gamma}\gamma n=1$, which is the case of $n=n_x$. Then we obtain
\begin{equation}
n_x\cong x_0^{-\gamma}~.
\label{a7}
\end{equation}

A comparison with third scaling hypothesis gives us that $z=-\gamma$. With
this procedure we obtained all the three critical exponents discussed in the
previous section as function of the parameter of the nonlinearity $\gamma$.
Numerical simulations were made for several different values of $\gamma$
considering either odd, even, irrational and other set of numbers. The
numerical findings confirm the validity of both the scaling law as well as the
analytical procedure.

The last point to discuss is the case of $R<R_c$, i.e., immediately before the
bifurcation. For this case we can rewrite the mapping as
\begin{eqnarray}
x_{n+1}-x_n&=&x_n(R-1)-Rx_n^{\gamma+1}~,\nonumber\\
&=&{{x_{n+1}-x_n}\over{(n+1)-n}}\approx{{dx}\over{dn}}~,\nonumber\\
&=&x(R-1)-Rx^{\gamma+1} ~.
\label{a8}
\end{eqnarray}

We have to emphasise that near the steady state $x\cong 0$ and considering
$\gamma>1$, the term $x^{\gamma+1}$ goes faster to zero as compared with $x$.
Then the last term of Eq. (\ref{a8}) can be disregarded. With this approach we
obtain the following differential equation
\begin{equation}
{{dx}\over{dn}}=-x\mu~,
\label{a9}
\end{equation}
where $\mu=1-R$. Considering again that for $n=0$ the initial condition is
$x_0$, we have to integrate the following equation
\begin{equation}
\int_{x_0}^{x(n)}{{dx}\over{x}}=-\mu\int_0^{n}dn^{\prime}~.
\label{a10}
\end{equation}
After integration and grouping the terms we obtain
\begin{equation}
x(n)=x_0e^{-\mu n}~.
\label{a11}
\end{equation}

Comparing this result with Eqs. (\ref{decay_exp}) and (\ref{decay_tau}) we
conclude that the exponent $\delta=-1$. This finding is in good agreement with
the simulations shown in Fig. \ref{Fig4}.
%*******************************************************

%***********************************************************
\section{Discussions}
\label{sec4}

\begin{figure}[t]
%\vspace*{-0.8cm}
\centerline{\includegraphics[width=1.0\linewidth]{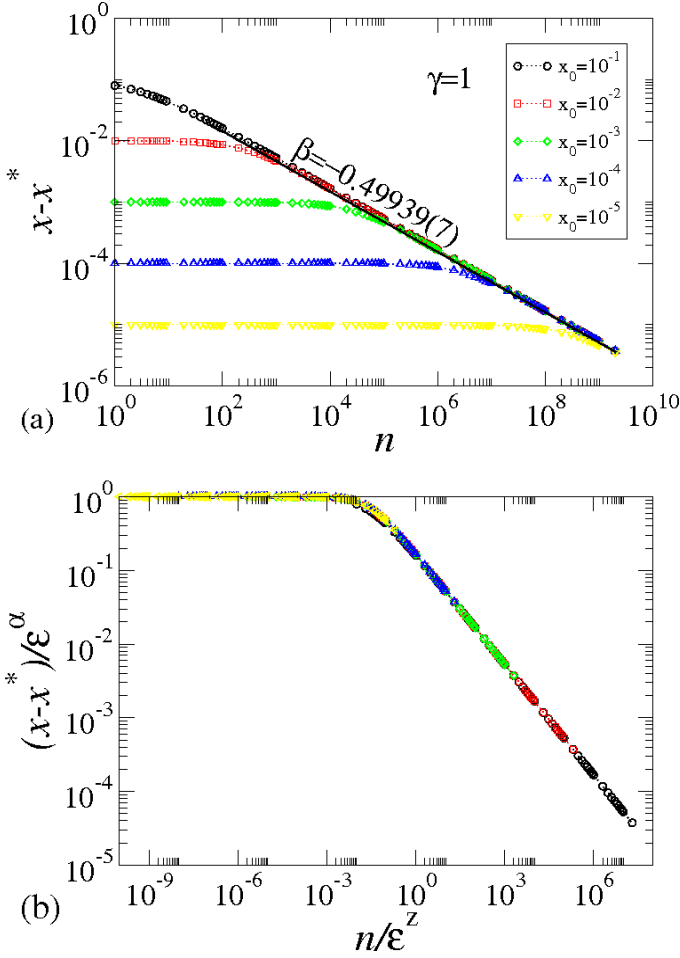}}
\caption{(a) Convergence to the fixed point $x^*=2/3$ for $\gamma=1$
considering $R=3$. The slope of decay obtained was $\beta=-0.49939(7)$. (b)
Overlap of the curves shown in (a) onto a single and hence universal plot.}
\label{Fig5}
\end{figure}

\begin{figure}[t]
%\vspace*{-0.8cm}
\centerline{\includegraphics[width=1.0\linewidth]{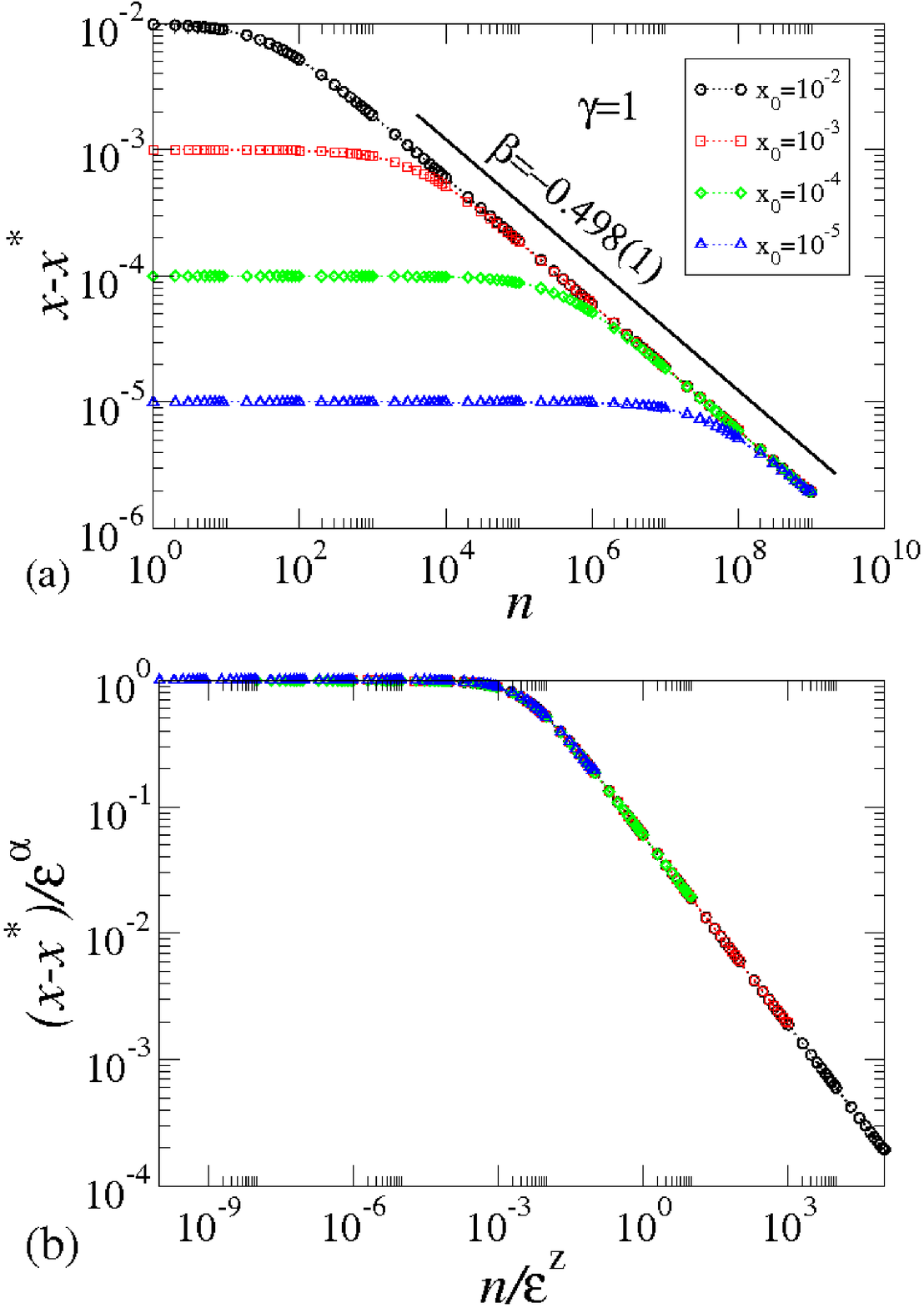}}
\caption{(a) Convergence to the fixed point $x^*_2=0.4399601688\ldots$ for
$\gamma=1$ considering $R=1+\sqrt{6}$. The slope of decay obtained was
$\beta=-0.498(1)$. (b) Overlap of the curves shown in (a) onto a single and
hence universal plot.}
\label{Fig6}
\end{figure}

Convergence to the steady state at a period-doubling bifurcation was also
observed to obey a homogeneous function. Indeed to apply the formalism we have
to look at the distance to the fixed point. Such observable can be defined as
(following Refs. \cite{Add1,Add2}) $y(n)=F^n(F^m(x))-x^*$ where $F$ stands for
the mapping, $m=2,4,6,\ldots$ and $x^*$ is indeed the expression of the fixed
point. In the previous sections, the fixed point was located at $x^*=0$,
hence it was convenient to rescale the observables in terms of $x_0$, which
was the initial distance from the fixed point. In the period-doubling
bifurcation considered here, $x^*$ is not zero anymore, hence we represent
the distance from the fixed point as $\epsilon$. Furthermore, the scaling
is dependent on $\epsilon$. Figure \ref{Fig5} shows the convergence to the
fixed point for $\gamma=1$ and $R=3$. The slope obtained for the decay is
$\beta=-0.49939(7)$. The crossover exponent was obtained as $z=-2.001(4)$,
and the same scaling law as obtained previously is applied here too. The
exponents obtained for the convergence towards the fixed point in the
period-doubling bifurcation show to be independent of the nonlinearity
$\gamma$. The scaling (see the overlap of the curves shown in Fig.
\ref{Fig5}(a) onto a single and universal plot as shown in Fig. \ref{Fig5}(b))
is better seen after the transformations: (i) $(x-x^*)\rightarrow
(x-x^*)/\epsilon^{\alpha}$ where $\epsilon$ stands for the distance of the
initial condition to the fixed point $x^*$ and; (ii) $n\rightarrow
n/\epsilon^z$. For $\gamma=1$, the first period-doubling bifurcation happens
at $R=3$, and the fixed point is $x^*=2/3$. For the second period-doubling,
$R=1+\sqrt{6}$ and the two fixed points are
\begin{equation}
x^*_{1,2}={{1}\over{2}}\left[1+{{1}\over{R}}\pm{{1}\over{R}}\sqrt{R^2-2R-3}
\right]~.
\label{fixed_points}
\end{equation}
For $R=1+\sqrt{6}$, the two fixed points assume the values
$x^*_1=0.8499377796\ldots$ and $x^*_2=0.4399601688\ldots$. In this cases, the
critical exponents are $\alpha=1$, $z=-2$ and $\beta=-0.498(1)$.

The numerical results obtained for the parameter $R=1+\sqrt{6}$ are shown in
Fig. \ref{Fig6}. The convergence to the fixed point is plotted in Fig.
\ref{Fig6}(a) for different values of initial conditions, as shown in the
figure while the overlap onto a single and universal plot is shown in Fig.
\ref{Fig6}(b). The slope of the decay obtained is the same one as obtained for
the first period-doubling bifurcation observed at $R=3$. This result confirms
the independence of the critical exponents $\alpha$, $\beta$ and $z$, on the
period-doubling bifurcation, as a function of $\gamma$.

%*******************************************************

%*********************************************************
\section{Conclusions}
\label{sec5}
To summarize, we have considered the convergence to the steady state in a
family of logistic-like mappings in a bifurcation point near a
transcritical, pitchfork, period-doubling bifurcation, and around its
neighbouring. At the bifurcation point we used a phenomenological description
to prove that decay to the fixed point is described by using a
homogeneous function with three critical exponents. The three critical
exponents are related between themselves via a scaling law of the type
$z=\alpha/\beta$. Near the bifurcation point the convergence to the fixed
point is given by an exponential decay and the relaxation time is described by
a power law of the type $\tau\propto \mu^{\delta}$. We found $\delta=-1$ and
is independent on the nonlinearity of the mapping. The results obtained by the
phenomenological approach is confirmed by an analytical description and
is valid for any $\gamma\ge 1$. The exponents obtained for the period doubling
bifurcation were $\alpha=1$, $\beta\cong -0.5$, $z=-2$ and are independent of
the parameter $\gamma$.
%*******************************************************

\section*{ACKNOWLEDGMENTS}

RMNT thanks to FUNCAP. FCG thanks to CNPq and PROPe/FUNDUNESP. RNCF thanks
to CNPq. JAO thanks to CNPq, PROPe/FUNDUNESP and FAPESP (2014/18672-8). EDL
acknowledges support from CNPq, FAPESP (2012/23688-5) and FUNDUNESP. The
authors thank Jo\~ao L. Menicuci for fruitful discussions.
%********************************************************

%*******************************************************

%********************************************************

\end{document}